\documentclass[prl,twocolumn,showpacs,amsmath,amssymb,superscriptaddress]{revtex4-1}

\usepackage{graphicx}
\usepackage{dcolumn}
\usepackage{bm}
\usepackage{color}

\usepackage{amsmath}
\usepackage{amssymb}
\usepackage{latexsym}
\usepackage{multirow}
\usepackage{xcolor}
\usepackage{hyperref}
 
\usepackage{textcomp}

\begin{document}

\title{
Correlation decoupling of Casimir interaction in an electrolyte driven by external electric fields}
\date{\today}
\author{Guangle Du}
\affiliation{ School of Physical Sciences, University of Chinese Academy of Sciences (UCAS), Beijing 100049, China.}
\author{David S. Dean}\email{david.dean@u-bordeaux.fr}
\affiliation{Universit\'e Bordeaux, CNRS, LOMA, UMR 5798, F-33400 Talence, France.}
\affiliation{ Kavli Institute for Theoretical Sciences, University of Chinese Academy of Sciences (UCAS), Beijing 100049, China.}
\author{Bing Miao}\email{bmiao@ucas.ac.cn}
\affiliation{ Center of Materials Science and Optoelectronics Engineering, College of Materials Science and Opto-Electronic Technology, University of Chinese Academy of Sciences (UCAS), Beijing 100049, China.}
\author{Rudolf Podgornik}\email{podgornikrudolf@ucas.ac.cn}
\affiliation{ School of Physical Sciences, University of Chinese Academy of Sciences (UCAS), Beijing 100049, China.}
\affiliation{Wenzhou Institute, University of Chinese Academy of Sciences, Wenzhou, Zhejiang 325001, China.}
\affiliation{Institute of Physics, Chinese Academy of Sciences, Beijing 100190, China.}

\begin{abstract} 
It is well established that the long range van der Waals or thermal Casimir interaction between two semi-infinite dielectrics separated by a distance $H$ is screened by an intervening electrolyte. Here we show how this interaction is modified when an electric field of strength $E$ is applied parallel to the dielectric boundaries, leading to a non-equilibrium steady state with a current. The presence of the field induces a long range thermal repulsive interaction, scaling just like the thermal Casimir interaction between dielectrics without the intervening electrolyte, {\em i.e.} as  $1/H^3$. At small $E$ the effect is of order $E^2$ while at large fields it saturates to an $E$ independent value. We explain the results in terms of a decoupling mechanism between the charge density fluctuations of cations and anions at large applied fields.
\end{abstract}

\maketitle

\noindent{\em Introduction--}
Fluctuation induced interactions are ubiquitous both from the 
fundamental as well as applied point of view and can arise  due to both quantum and thermal fluctuations~\cite{par05,woo16,dan23}. These interactions can be tuned by changing material properties for van der Waals interactions~\cite{woo16} or by chemical variations of surface properties for the critical Casimir effect~\cite{RevModPhys.90.045001}. Another way to modify these interactions is to apply external fields, for example electric or magnetic fields~\cite{dea16,mah21}. 
The application of an electric field generates a current, in conducting systems, and this means that the system is out of equilibrium. The out of equilibrium nature of the problem means that standard equilibrium methods, {\sl e.g.} the Lifshitz formulation of van der Waals forces cannot be applied directly~\cite{Advances}, it  can however  be extended to compute van der Waals forces between dielectrics held at different temperatures~\cite{Antezza2, Antezza1,Kardar}. Out of equilibrium thermal or classical fluctuation induced forces can be studied via the Langevin dynamics of dipole fields in dielectrics and/or ionic Brownian dynamics in electrolytes~\cite{dea12,dea14,lu15,dea16}. 
\begin{figure}[!t]
  \centering
  \includegraphics[width=.9\linewidth]{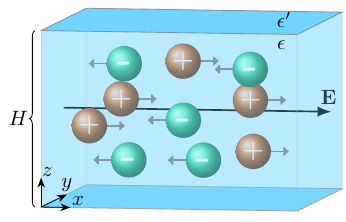}
  \caption{\label{fig:schematic} Electrolyte driven by an external electric field ${\bf E}$, in solvent of dielectric constant $\epsilon$, confined between two dielectric semispaces of dielectric constant $\epsilon'$. $H$ is the separation between the dielectrics, $z$ is the normal coordinate with the bottom surface located at $z=0$.}
\end{figure}
An illuminating model of the thermal Casimir effect is the {\sl living conductor model} where two plates containing an electrolyte interact across a dielectric gap~\cite{lev99,jan05}. In equilibrium this model exhibits the universal long range attractive Casimir interaction for free scalar field theories~\cite{Advances}. Recently~\cite{dea16}, it was shown that when an electric field is applied parallel to one plate this attractive Casimir interaction is reduced by the fact that the driving electric field destroys, or scrambles, the charge-charge correlations which yield the attraction in the equilibrium case. More recently~\cite{mah21}, the force between two planar dielectrics with an electrolyte in between, and driven by an electric field was studied as a function of the field strength (see the schematic in Fig.~(\ref{fig:schematic})). Here, it was found that driving the system leads to an effective long range interaction which can be attractive or repulsive. This is an intriguing result as, in equilibrium, the long range van der Waals interaction between two dielectric slabs with an intervening electrolyte is established to be screened, {\sl i.e.}, it decays exponentially with the plate separation at distances larger than the Debye screening length~\cite{nin76,net01}. The Casimir interaction in the presence of a driving electrostatic field was analyzed by using an approximation scheme for the linearized stochastic density functional  theory (SDFT) that describes the classical charge fluctuations in the system~\cite{mah21}. This leads to an effective theory reminiscent of statistical field theories where the introduction of anisotropy is mismatched with the thermal noise leading to long-range correlations~\cite{gar90,gri90} which in turn generate a long-range fluctuation induced interaction~\cite{Sengers1,Sengers2}. At small  $E$ the results of~\cite{mah21} predict a  force scaling as $E^4$, while at large fields it scales as $E^2$. These results are  surprising, first one might expect that the small field force should scale as $E^2$ unless there is some fine tuning arising naturally in the problem, secondly it is not clear how a laterally imposed large electric field could lead to arbitrarily large forces.

{ Here we reconsider the problem studied in Ref.~\cite{mah21} in the limit where the local dielectric constant of the surrounding (outer) dielectrics  $\epsilon({\bf x})= \epsilon'$ is much smaller than the dielectric constant of the intervening solvent where  $\epsilon({\bf x})=\epsilon$}. This case can be solved {\sl without any approximation} beyond the linearization of the SDFT for the electrolyte.  
We find that an electric field does indeed induce a long range interaction  and this interaction gives  a total, or net, {\sl repulsive force} per unit area 
\begin{equation}
f_{\rm t}(H)= \frac{\zeta(3)}{8\pi \beta H^3}\left[1- \left({1+\frac{\beta^2q^2 E^2 }{\kappa^2}}\right)^{-{\textstyle\frac12}}\right].\label{mainasym}
\end{equation}
as well as exponentially screened terms which are thus subdominant for large $H$. In Eq.~(\ref{mainasym}) $\beta=1/(k_{\mathrm{B}}T)$ where  $k_{\mathrm{B}}$ is Boltzmann's constant and $T$ is the temperature of the system (assumed constant), $\pm q$ are the charges of the cations/anions, and $\kappa = \sqrt{2\overline \rho q^2/(\epsilon k_{\mathrm{B}}T )}$ is the inverse Debye screening length where $\overline\rho$ is the same mean density of cations and anions. { Notice that the $E$ dependence of the long range force occurs via the coupling constant $g= \beta E q/\kappa$ which can be written as $g= v/v^*$, where $v=\beta D E q$  is the absolute drift velocity of ions due to the external   electric field and $v^*=D\kappa$ is an intrinsic velocity scale for the non-driven electrolyte system, corresponding exactly to the same non-equilibrium parameter found in Ref.~\cite{dea16} for the Casimir interaction between two plates containing Brownian  electrolytes,  where one is driven by an electric field. Only the thermal component of the van der Waals interaction is modified by the driving of the ions by the field. However, in aqueous systems this thermal component is typically up to $50\%$ of the total interaction \cite{woo16}. At room temperature the coupling constant $g$ can be rewritten as $g \approx E/E_0$ where $E_0 = V_0/\lambda_D$ is the thermal field with $V_0$ being the thermal voltage, $V_0 = k_BT/q  \sim 25 mV$, and $\lambda_D$ the Debye length, thus $E_0 \sim 2.5 \times 10^{6} V/m$ for a $10 nm$ Debye length. Assuming then a $\sim 30 V$ potential difference for electrodes $\sim 10 \mu m$ apart \cite{Perkin2019}, we have $g \sim 1$ and $E$ still below the water dissociation value  $\sim 10^{8} V/m$.}


The {\sl long range force} in this dielectric set up is always repulsive, scales as $E^2$ for small $E$ and saturates for large applied fields. We also identify  the origin of this limiting form of the repulsion at large electric fields which is due to the decoupling of the cationic and anionic  charge fluctuations, which then become independent of one another for large fields. For $E=0$, we recover the well known result for the screened thermal Casimir interaction~\cite{nin76,net01}, from our purely dynamical approach.

\noindent{\em The model--}
Consider (see Fig.~(\ref{fig:schematic})) two semi-infinite dielectrics of dielectric constant $\epsilon'$ separated by a slab of thickness $H$ in the direction $z$, containing an electrolyte solution of dielectric constant $\epsilon$ and 
ionic densities $\overline \rho$ of cations and anions of charges $\pm q$, but with otherwise identical properties, notably they have the same friction coefficient $\gamma$. These conditions of symmetry can be relaxed but the resulting formulas become much more complicated. This symmetric case is sufficient to highlight the underlying physics of the problem.  The over-damped Langevin equation for the anions and cations (labeled by the index $i$) is given by
\begin{equation}
\gamma \frac{\mathrm{d} {\bf x}_i}{\mathrm{d}t}=    q_i({\bf E}-\nabla\phi({\bf x}_i)) +\sqrt{2k_{\mathrm{B}}T\gamma}~\bm{\eta}_i(t),
\end{equation}
which corresponds to force balance between the frictional force on the left hand side and the electrical and thermal noise  forces on the right hand side. The first term on the right is the force due to the applied electric field parallel to the slabs (taken to be in the direction $x$) ${\bf E}= E {\bf e}_x$, and the second one is the force due to the electric field generated by the ions themselves:
\begin{equation}
\phi({\bf x}) = q\int \mathrm{d}{\bf x}' G({\bf x},{\bf x}') [\rho_+({\bf x}')-\rho_-({\bf x}')].\label{phi}
\end{equation}
In Eq.~(\ref{phi}) $\rho_{\pm}({\bf x}) = \sum_{i\pm}
\delta({\bf x}-{\bf x}_i)$ denotes the density of cations/anions  and $G({\bf x},{\bf x}')$ is the Green's function for the slab geometry where
\begin{equation}
\nabla\cdot \epsilon({\bf x}) \nabla G({\bf x},{\bf x}') = -\delta({\bf x}-{\bf x}').\label{gf}
\end{equation}
Finally, the term $\bm{\eta}_i(t)$ denotes a Gaussian white noise of zero mean with $\langle \eta_{i\alpha}(t)\eta_{j\beta} (t')\rangle = \delta_{ij}\delta_{\alpha\beta}\delta(t-t')$, where $\alpha$ and $\beta$ label the spatial components $(x,\,y,\,z)$. Starting from the Langevin equations we can formally write down the SDFT for the densities $\rho_\pm({\bf x})$ \cite{dea96,dem16}, this non-linear theory  becomes solvable if we expand about a background of constant density $\rho_\pm({\bf x})=\overline\rho +n_{\pm}({\bf x})$. Note that when the electric field is parallel to the bounding surfaces it does not generate a surface charge and so the expansion of the density about its average bulk value is justified. The analysis of the linearized SDFT is simplified by using the fluctuations of the total density $n_{\mathrm{t}}({\bf x})  = n_+({\bf x}) + n_-({\bf x})$ and the fluctuations of the charge density 
$\rho_{\mathrm{c}}({\bf x}) = q[n_+({\bf x}) - n_-({\bf x})] = q n_{\mathrm{d}}({\bf x})$. One thus finds~\cite{SM}
\begin{eqnarray}
\frac{\partial n_{\mathrm{t}}({\bf x})}{\partial t} &=& \nabla\cdot D[\nabla n_{\mathrm{t}}({\bf x}) - \beta q n_{\mathrm{d}}({\bf x}) {\bf E}] + \xi_{\mathrm{t}}({\bf x},t),
\label{sd1}\\
\frac{\partial n_{\mathrm{d}}({\bf x})}{\partial t} &=& \nabla\cdot D[\nabla n_{\mathrm{d}}({\bf x}) -\beta q n_{\mathrm{t}}({\bf x}) {\bf E} + 2q\beta\overline\rho\nabla \phi({\bf x})] \nonumber\\&& + \xi_{\mathrm{d}}({\bf x},t).\label{sd2}
\end{eqnarray}
where the spatio-temporal white noise terms are independent and have correlation functions $\langle\xi_{\mu}({\bf x},t)\xi_{\nu}({\bf x}',t')\rangle= -4\overline\rho D\delta_{\mu\nu}\delta(t-t') \nabla^2\delta({\bf x}-{\bf x}'),\,\mu,\nu\in\{\mathrm{t},\mathrm{d}\}$. The term $D$ is the diffusion constant of the cations and anions, given by $D =k_{\mathrm{B}}T/\gamma$. The boundary conditions on the linearized SDFT are no-flux boundary conditions, so $\nabla_z n_{\mathrm{t}}({\bf x})=0$ and $\nabla_z n_{\mathrm{d}}({\bf x})+ 2q\beta\overline\rho\nabla_z \phi({\bf x})=0$ at the boundaries $z=0$ and $z=H$. No  flux boundary conditions are also imposed independently on the noise terms, as the noise is independent of the ionic distribution at the time when it acts. The boundary condition on $n_{\mathrm{d}}({\bf x})$  is non-local due to the appearance of the electrostatic potential which depends on the whole charge density in the problem.  This non-locality can be dealt with~\cite{Future}, but here we consider the simplifying case where we assume that $\epsilon'\ll \epsilon$. This means that both the density and the electrostatic potential have the same, Neumann, boundary conditions at the bounding surfaces. 
Consequently,  both fields, $n_{\mathrm{t}}({\bf x})$ and $n_{\mathrm{d}}({\bf x})$, as well as the noise fields, can be expanded in terms of a Fourier cosine expansion~\cite{SM}, yielding 
\begin{widetext}
\begin{eqnarray}
n_{\mu}({\bf x}_{\|},z,t) &=&\int \frac{\mathrm{d}{\bf k}}{(2\pi)^2} \sum_{n=0}^{\infty} \frac{1}{\sqrt{N_n}}\tilde n_{\mu n}({\bf k},t)\exp(\mathrm{i}{\bf k}\cdot {\bf x}_{\|})\cos(p_n z),\, \mu \in \{\mathrm{t}, \mathrm{d}\},
\end{eqnarray}
\end{widetext}
where $p_n = n\pi/H$ enforces the Neumann conditions at $z=0$ and $z=H$, while $N_n$ normalizes the corresponding eigenfunction, with $N_n = H/2$ for $n\geq 1$ and $N_0=H$. The Fourier transform is taken in the plane $\mathbf{x}_{\|}=(x,y)$ parallel the bounding dielectric surfaces. In the steady state the resulting Lyapunov equation~\cite{SM} can be solved analytically for the equal time correlation functions  
$\langle \tilde n_{\mu m}({\bf k},t)\tilde n_{\nu n}({\bf k}',t)\rangle = (2\pi)^2 \delta_{mn}\delta({\bf k}+{\bf k'}) \tilde C_{\mu\nu n}({\bf k})$, $\mu,\nu\in \{\mathrm{t},\mathrm{d}\}$,
where
\begin{widetext}
\begin{subequations}
\begin{eqnarray}
\tilde C_{\mathrm{tt}n}({\bf k}) &=& 2\overline\rho \frac{(k^2 + p_n^2) \left[ (K^2 + p_n^2) (K^2 + k^2 + 2 p_n^2) + 2 \beta^2 q^2 E^2 k_x^2  \right]}{(K^2 + k^2 + 2 p_n^2)\left[(K^2 + p_n^2)(k^2 + p_n^2) + \beta^2 q^2 E^2 k_x^2  \right]}, \\
\tilde C_{\mathrm{td}n}({\bf k}) &=& -\tilde C_{\mathrm{dt}n}({\bf k})= \mathrm{i} 2\overline \rho \frac{\beta q E k_x \kappa^2 (k^2 + p_n^2)}{(K^2 + k^2 + 2 p_n^2)\left[(K^2 + p_n^2)(k^2 + p_n^2) + \beta^2 q^2 E^2 k_x^2  \right]}, \label{eq:corrCationAnion}\\
\tilde C_{\mathrm{dd}n}({\bf k}) &=&  2\overline\rho \frac{(k^2 + p_n^2) \left[ (k^2 + p_n^2) (K^2 + k^2 + 2 p_n^2) + 2 \beta^2 q^2 E^2 k_x^2  \right]}{(K^2 + k^2 + 2 p_n^2)\left[(K^2 + p_n^2)(k^2 + p_n^2) + \beta^2 q^2 E^2 k_x^2  \right]}
\end{eqnarray}
\end{subequations}
\end{widetext}
with $K^2 = k^2 +\kappa^2$ and $k_x={\bf e}_x\cdot{\bf k}$.

We now turn to the computation of the force between the two dielectrics. In writing Eq.~(\ref{gf}) we have implicitly assumed that the dipole degrees of freedom in the system adjust {\sl instantaneously} to generate image charges, and are thus fully equilibrated with respect to a given configuration of charge density (dynamics of dipoles being governed by rapid electronic degrees of freedom  orders of magnitude faster than the diffusive dynamics of the ions of the electrolyte). However, in principle, potentially slower dynamics of the dipoles in the dielectrics can be taken into account~\cite{dea12}.  
The dipole fluctuations  also  generate their own thermal component of the Casimir force between two dielectrics, which would  be present in the absence of any electrolyte, {\em i.e.}, when $\overline\rho=0$. There is also an ideal gas component coming from the ions when they are uncharged or due to the constant non-fluctuating part of the densities which is by definition also assumed to be equilibrated. This means that the average total force per unit area is given by
\begin{equation}
f_{\rm t}= f_{\rm vdW} + f_{\rm ion}+ 2\,\overline{\rho}\,k_{\mathrm{B}}T ,\label{feff}
\end{equation}
where $f_{\rm vdW}$ is  the van der Waals interaction between the two dielectrics in the absence of ions  (including contributions from the zero and non-zero Matsubara frequencies), which are assumed to be instantaneously equilibrated, and $2\,\overline{\rho}\,k_{\mathrm{B}}T$ is the bulk ideal van't Hoff pressure term, which is also assumed to be equilibrated as it is not affected by the external field, and will be ignored subsequently since it is independent of separation $H$. The second term $f_{\rm ion}$ is the contribution due to the ions which can be deduced by constructing a stress tensor from the local body force, which can then be written down in terms of the density fluctuations, and is given by~\cite{SM,kru18}
\begin{equation}
\begin{aligned}
\sigma_{\alpha\beta}({\bf x})=&-\frac{k_{\mathrm{B}}T}{4\,\overline \rho}\delta_{\alpha\beta}(n^2_{\mathrm{t}}({\bf x})+n^2_{\mathrm{d}}({\bf x})) + \\
&\epsilon({\bf x})(\nabla_\alpha\phi({\bf x})\nabla_\beta\phi({\bf x})-\frac{1}{2}\delta_{\alpha\beta}[\nabla\phi({\bf x})]^2).
\end{aligned}
\end{equation}
The first term on the right hand side is generated by the density fluctuations, while the second term is the contribution from the standard Maxwell stress tensor. Furthermore, in the limit $\epsilon'\ll \epsilon$ only the part of the stress tensor inside the electrolyte slab yields a non-zero contribution to the force.

 The outward acting surface force density due to the ions at $z=0$ is given by  $f_{\rm ion}= -\langle\sigma_{z\,z}(0, 0, 0^+)-\sigma_{z\,z}(0, 0, 0^-)\rangle$  (by translational invariance in the $(x,y)$ plane). The average is taken over the spatio-temporal noise, the minus sign comes from the downward normal and $z = 0^\pm$ denotes the position just above and just below the bounding surface at $z=0$, respectively. In the limit $\epsilon' \ll \epsilon$  the stress tensor is diagonal and we find
\begin{equation}
f_{\rm ion}=\frac{k_{\mathrm{B}}T}{4\overline \rho}
\langle n^2_{\mathrm{t}}({\bf 0})+n^2_{\mathrm{d}}({\bf 0})\rangle +\frac{ \epsilon}{2}\langle[\nabla_{\|}\phi({\bf 0} )]^2\rangle,
\end{equation}
where $\nabla_{\|}$ denotes the gradient in the $(x,y)$ plane. 
One can explicitly extract the terms which give the total $H$ dependent force (including the contribution from the dielectric slabs $f_{\rm vdW}$ in Eq.~(\ref{feff}))~\cite{SM}, these are
\begin{widetext}
\begin{equation}
f_{\rm t}(H)= 
-\frac{1}{\beta} \int \frac{\mathrm{d}{\bf k}}{(2\pi)^2}\sum_n \frac{1}{2 N_n} \frac{K^2(k^2 + p_n^2)^2(K^2+k^2 + 2 p_n^2) + \beta^2 q^2 E^2 k_x^2 \left[2k^2(k^2 + p_n^2) + \kappa^2 (k^2 + 2 p_n^2)\right]}{(k^2 + p_n^2)(K^2 + k^2 + 2 p_n^2)\left[(k^2 + p_n^2)(K^2 + p_n^2) + \beta^2 q^2 E^2 k_x^2 \right]}.\label{fH}
\end{equation}
\end{widetext}

\begin{figure}[tb!]
\centering
  \includegraphics[width=.99\linewidth]{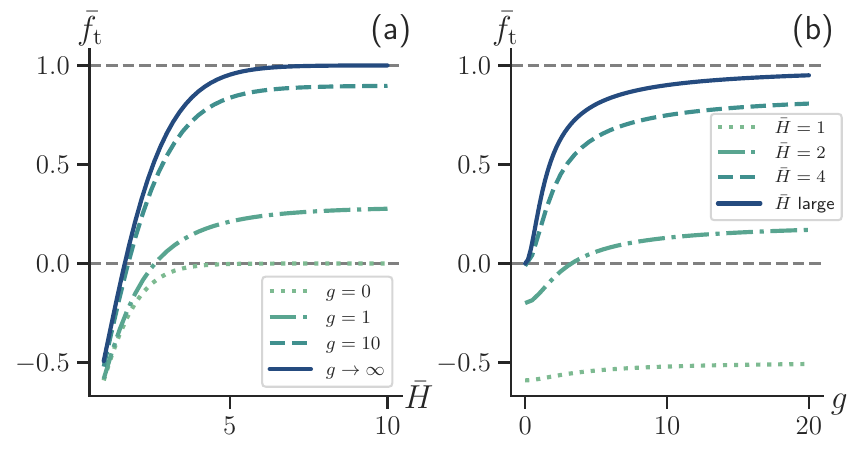}
\caption{\label{fig:graph} Dependence of the interaction force $\bar{f}_{\mathrm{t}}=f_{\mathrm{t}}/[\zeta(3)/8\pi \beta H^3]$ on (a) the separation $\bar{H} = H \kappa$ with fixed coupling constant $g = \beta q E /\kappa$ and (b) $g$ with fixed $\bar{H}$. The asymptotic large $\bar{H}$ and $g$ results are shown as the dashed lines $\bar{f}_{\mathrm{t}}=1$.}
\end{figure}

For  $E=0$, the above, purely dynamically derived, equation yields the standard equilibrium result for the screened thermal Casimir interaction which can be written in the same limit $\epsilon'\ll \epsilon $ as~\cite{nin76} 
\begin{eqnarray}
f_{\rm t}(H) &=& -\frac{1}{4\pi \beta H^3} \int_{\kappa H}^{\infty} \mathrm{d} K \,K^2\left(\coth K - 1 \right). 
\label{tsc}
\end{eqnarray}
The charge fluctuations of the electrolyte obviously produce a repulsive thermal Casimir interaction which cancels {\sl exactly} the attractive thermal Casimir interaction between the two bounding dielectrics. This effect was already pointed out explicitly for an equilibrium system in~\cite{jan04} and was also discussed previously by a number of other authors~\cite{att87,att88,dea02}. In the $E=0$ limiting case this attractive interaction is 
\begin{equation} \label{eq:vdwForce}
f_{\rm vdW}(H) = - \frac{\zeta(3)}{8\pi\beta H^3}.
\end{equation}
The residual interaction after this cancellation then turns out to be attractive and screened. 

The integral and sum in Eq.~(\ref{fH}) can be evaluated analytically~\cite{SM} but the result is rather unwieldy and does not have an intuitive interpretation. However, the large distance behavior of the interaction, more precisely in the limit where $H\kappa\gg1$, can be simply extracted from the small $k$ and $p_n$ components in Eq.~(\ref{fH}). This then gives for large $H$ the particularly simple formula in Eq.~(\ref{mainasym}) for the dominant large distance component of
$f_{\rm t}(H)$, which is exact up to terms which decay as $\exp(-2\kappa H)$. We emphasize that the long distance repulsion is only present for non zero $E$ and is a purely non-equilibrium effect. For small $E$ the force behaves as
$f_{\rm t}(H)= \frac{\zeta(3)}{16\pi\beta H^3}\frac{\beta^2 E^2 q^2}{\kappa^2}$.
On the other hand, for strong electric field driving as $E\to\infty$ in the same limit of $\epsilon' \ll \epsilon$,   we find 
\begin{equation}
f_{\rm t}(H)\!=\! \frac{\zeta(3)}{8\pi\beta H^3}- \frac{2}{4\pi\beta H^3}\!\!\int_{\frac{\kappa H}{\sqrt{2}}}^{\infty}\!\!\mathrm{d}K K^2\!\left(\coth K - 1 \right).\label{rep}
\end{equation}
The first term is a {\sl repulsive} thermal Casimir force of exactly the same absolute magnitude as the equilibrium {\sl attraction} between two dielectrics with no intervening electrolyte. This result can be understood as a correlation decoupling between the cations and anions due to the strong external electric field~\cite{SM}, indicated by Eq.~(\ref{eq:corrCationAnion}). If only the cationic type was present then the effect of the electric field would be to drive all the cations at the same average velocity $v=\beta D q E$. In the rest frame of the cations the system would then be in an effective equilibrium. Charge fluctuations in this equilibrium state would then generate a repulsive thermal Casimir interaction as given by Eq.~(\ref{rep}). However,  the same argument is equally valid for the anions that move with velocity $v=-\beta D q E$. The crucial point is that  the dynamics of the cations and anions becomes {\sl decoupled} in the limit of $E\to\infty$, and so they both generate 
a repulsive force of the form in Eq.~(\ref{rep}), subtracting off the attractive thermal Casimir interaction of the dielectrics then giving the net repulsive force Eq.~(\ref{rep}).  

The second term in Eq.~(\ref{rep}), which is screened, is also consistent with the above interpretation in terms of the correlation decoupling~\cite{SM}. In fact the inverse screening length of $\kappa/\sqrt{2}$ indicates that the ions of either type are screening independently. Since in the rest frame of the anions/cations they are in an effective equilibrium at half the total density (see above), the Casimir interaction is exactly twice (explicit factor 2 in the second term of Eq.~(\ref{rep})) the screened Casimir force at $\kappa/\sqrt{2}$ screening length, giving the strong electric field driving limiting law a consistent intuitive interpretation. 

The full force $f_{\rm t}$ at arbitrary  $H$ and and $E$ can be written in  terms of integrals which need to be evaluated numerically \cite{SM}. In Fig.~(\ref{fig:graph}) is shown $f_{\rm t}$  as a function of $\bar H= H\kappa$ ({\em i.e.} with $H$ measured in units of the Debye length) and $g$, where $f_t$ are normalized by the absolute value of van der Waals attraction in Eq.~(\ref{eq:vdwForce}). In Fig.~(\ref{fig:graph}a) we see the behavior of $f_{\rm t}$ at fixed $g$ as $\bar H $ is varied. The lowest curve corresponds to $g=0$ the screened thermal Casimir force given in Eq.~(\ref{tsc}) which is always attractive. The cases $g=1,\ 10$ and $g\to\infty$ are also shown, at short distances all these forces are attractive but on increasing $\bar H$ they become repulsive, the crossover value of $\bar H$ from attraction to repulsion decreases on increasing $g$. The large $g$ limits agree with the asymptotic result Eq.~(\ref{rep}). In Fig (\ref{fig:graph}b) we show $f_{\rm{t}}$ at fixed values of $\bar H$ as a function of $g$.
At all separations increasing $g$ increases the repulsive component of the force, at short distances the force remains net attractive but at large distances it eventually becomes net repulsive. The large $\bar{H}$ limits agree with the asymptotic result Eq.~(\ref{mainasym}). Furthermore,  $f_t$ saturates to the repulsion $\zeta(3)/(8\pi\beta H^3)$ at either large separation $H$ or large coupling $g$.

\noindent{\em Conclusions--} {Using non-equilibrium SDFT  for an electrolyte confined between two dielectrics in an external driving electric field, we first of all recover the screened thermal Casimir interaction from a purely dynamical approach. For asymptotically large separations the behavior is given by the simple form of Eq.(\ref{mainasym}). At large $E$ exponentially screened terms correspond to the standard screened thermal Casimir force with the Debye screening length for vanishing electric field, and to a Debye screening length at half the ion density for large electric fields, this effect being due to decoupled ionic correlations in the presence of strong external driving. Many extensions of this work are possible such as the extension to non-symmetric electrolytes (both charges and ion mobilities) and to the case where the field is applied perpendicular to the surfaces which requires the application of SDFT to inhomogeneous systems.
The non-equilibrium effects predicted here could  be tested either in the surface force apparatus~\cite{ric20} or colloidal probe atomic force microscopy interaction geometry~\cite{Trefalt20}.} 

\noindent{\em Acknowledgments--}
The authors would like to thank Carlo Drummond for useful discussions. G.D., B.M., and R.P. acknowledge funding from the Key Project No. 12034019 of the National Natural Science Foundation of China and the support by the Fundamental Research Funds for the Central Universities (Grant No. E2EG0204). D.S.D. acknowledges support  from the grant No. ANR-23-CE30-0020-01 EDIPS, and by the European Union through the European Research Council by the EMet-Brown (ERC-CoG-101039103) grant.

\newpage
\onecolumngrid
\appendix
\renewcommand{\theequation}{S\arabic{equation}}
\setcounter{equation}{0}
\section{Supplementary Material}
This Supplementary Material  provides additional  technical details of the physical model and of the calculations in the letter. 

{\em Section (I)--} First we discuss the model for the confined electrolyte  and how the fast relaxation of the dielectric components of the model with respect to the slow ionic degrees of freedom lead to an effective model of the ions where image charges are introduced as well as a thermal  van der Waals interaction due to the direct interaction of the dielectric materials in the system.  

{\em Section (II)--} We then carry out a Debye-H\"uckel type of analysis where we expand the density fluctuations to second order about the mean density, deriving an effective quadratic  Hamiltonian for the density fluctuations. For completeness sake, within this formalism, we rederive the well known equilibrium results for the screened  thermal Casimir interaction between two dielectric slabs sandwiching an electrolyte in the limit where $\epsilon'$, the dielectric constant of the surrounding dielectric medium  is much smaller than $\epsilon$ the dielectric constant of the intervening solvent which contains the electrolyte.

{\em Section (III)--} We derive the stress tensor to compute the forces generated by the ionic degrees of freedom for a general non-equilibrium situation and show how ionic density fluctuations lead to an additional  term in the stress tensor with respect to the Maxwell stress tensor. 

{\em Section (IV)--} We write down  the dynamical equations for the densities of the anions and cations and
show how the linearized form of these equations leads to a model B type conserved dynamics with the quadratic Hamiltonian of Section (2). We solve these equations in the presence of a driving electric field ${\bf E}= E{\bf e}_x$ parallel to the surfaces, here in the $x$ direction,  to compute the correlation functions of the density fluctuations in the resulting steady state.

{\em Section (V)--} We use the results of Section (4) to compute the $H$ (separation between the two surfaces bounding the dielectric) dependent part of the interaction between the two slabs. We show how in the case where $E=0$ we recover the equilibrium screened thermal Casimir force. We derive a full integral expression for the $H$ dependent force between the two slabs. We then derive the explicit formula for the long range (non-screened) component of these forces.

\section{Effective Hamiltonian  for the ions }
Here we discuss what is referred to in Ref.~\cite{jan04} as the {\sl separation hypothesis} which gives an effective  electrostatic Hamiltonian  for the ions,  including the image charges, as well as introducing an effective unscreened van der Waals interaction between two dielectric slabs, independent of the charge density between the slabs (and is indeed present in the absence of charge density). 

The partition function for a fixed charge density in a medium with a local dipole field ${\bf p}({\bf x})$ and local polarizability $\chi({\bf x})$ can be written as
\begin{equation}
Z=\int \mathrm{d}[\phi] \mathrm{d}[{\bf p}] \exp\left(-\frac{1}{2}\beta\epsilon_0 \int \mathrm{d}{\bf x} [\nabla\phi({\bf x}) ]^2 + \mathrm{i} \beta \int \mathrm{d}{\bf x}
\phi({\bf x}) [\rho_{\mathrm{ic}}({\bf x})- \nabla\cdot {\bf p}({\bf x}) ] -\frac{1}{2}\beta\int \mathrm{d}{\bf x}\frac{ {\bf p}^2({\bf x})}{\chi({\bf x})}\right)
\end{equation}
The total charge density in the above is $\rho_c({\bf x}) = \rho_{\mathrm{ic}}({\bf x})- \nabla\cdot {\bf p}({\bf x})$, the first term being the ionic charge density and the second the charge density due to the dipoles in the dielectrics and solvent. Integrating over the field $\phi$ simply introduces the direct pure Coulomb interaction between the total charge densities in the problem and the determinant involved in the integration does not depend on the configuration of the charge densities.

The local quadratic term  $\frac{ {\bf p}^2({\bf x})}{2\chi({\bf x})}$ corresponds to a simple harmonic spring model for the local dipole field, but at the level of linear response other models are equivalent. 
We will see below how $\chi({\bf x})$ is related to the local static dielectric constant.
The above model can be used to study the out of equilibrium behavior of the thermal and/or non-zero Matsubara contribution to the van der Waals forces. However, here we assume that the dynamics of the field 
${\bf p}({\bf x})$ is so rapid that it equilibrates instantaneously with any charge density configuration of the ions.
Integrating over the field ${\bf p}$ then yields (up to an overall prefactor)
\begin{equation}
Z=\int \mathrm{d}[\phi] \exp\left(-\frac{1}{2}\beta \int \mathrm{d}{\bf x} \epsilon({\bf x}) [\nabla\phi({\bf x}) ]^2 +\mathrm{i}\beta \int \mathrm{d}{\bf x}
\phi({\bf x}) \rho_{\rm ic}({\bf x})\right),
\end{equation}
where $\epsilon({\bf x})= \epsilon_0 +\chi({\bf x})$ can be simply interpreted as the local dielectric constant.
Strictly speaking the above integration over the dipole field introduces a bulk term in the free energy but this term does not give an $H$ dependent contribution to  the force, so we will ignore it for the purposes of our analysis. Now carrying out the Gaussian integration over the field $\phi$ we find
\begin{equation}
Z = Z_{\rm vdW}\exp(-\beta H_{\rm ES}[\rho_{\rm ic}]),
\end{equation}
where $H_{\rm ES}$ is the electrostatic Hamiltonian 
\begin{equation}
H_{\rm ES}[\rho_{\rm ic}]=\frac{1}{2}\int \mathrm{d}{\bf x} \mathrm{d}{\bf x}' \rho_{\rm ic}({\bf x})G({\bf x},{\bf x}')\rho_{\rm ic}({\bf x}')
\end{equation}
with $G({\bf x},{\bf x}')$ the Green's function
\begin{equation}
\nabla\cdot \epsilon({\bf x})\nabla G({\bf x},{\bf x}')=-\delta({\bf x}-{\bf x}'), \label{gf}
\end{equation}
and 
\begin{equation}
 Z_{\rm vdW}=\int \mathrm{d}[\phi]  \exp\left(-\frac{1}{2}\beta \int \mathrm{d}{\bf x} \epsilon({\bf x}) [\nabla\phi({\bf x}) ]^2 \right)\label{zvdw}
\end{equation}
the partition function for the van der Waals interaction in the system in the absence of any ionic charge density. At a risk of laboring the point therefore we see that assuming that the dipole degrees of freedom equilibrate instantaneously for any fixed ionic charge distribution, the effective interaction between the ionic charges is mediated by the Green's function $G$ in Eq.~(\ref{gf}) which contains the effects of image charges. There is also a direct van der Waals free energy, corresponding to $Z_{\rm vdW}$, which would be present even without any electrolyte. There are many ways to compute $Z_{\rm vdW}$ and the result is well known. However for the sake of completeness we will calculate  it out here using an eigenfunction expansion method as this method will be used to obtain results for the driven system.

We can write  the field $\phi$  as a Fourier/eigenfunction expansion 
\begin{equation}
\phi(x,y,z)= \phi({\bf x}_{\|},z) = \int \frac{\mathrm{d}{\bf k}}{(2\pi)^2}\sum_{n} \tilde{\phi}_n({\bf k})\psi_n(z)\exp(\mathrm{i}{\bf k}\cdot{\bf x}_{\|}),
\end{equation} 
with boundary condition
\begin{equation}
\epsilon'\left. \frac{\partial \psi_n}{\partial z}\right|_{z=0^-,H^+} = \epsilon \left.\frac{\partial \psi_n}{\partial z}\right|_{z=0^+,H^-}
\end{equation}
at $z=0$ and $z=H$. This means that when $\epsilon\gg\epsilon'$ 
the boundary conditions at $z=0,H$ become 
\begin{equation}
\left.\frac{\partial \psi_n}{\partial z}\right|_{z=0^+,H^-} = 0
\end{equation}
and so 
\begin{equation}
\psi_0(z) = \frac{1}{\sqrt{H}} ,\ {\rm and} \ \psi_n(z)=\frac{\sqrt{2}}{\sqrt{H}}\cos(p_n z)\ {\rm for}\ n\ge 1,
\end{equation}
where $p_n = n\pi/H$.
Using this gives a product of simple Gaussian integrals with each mode contributing an $H$ dependent factor
\begin{equation}
Z_{\mathrm{vdW},n,{\bf k}}= \left( k^2 + p_n^2\right)^{-\frac{1}{2}}.
\end{equation}
Computing the van der Waals force per unit area from the resulting free energy we then find
\begin{equation}
f_{\rm vdW} = \frac{k_{\mathrm{B}}T}{(2\pi)^2H}\int \mathrm{d}{\bf k}\sideset{}{'}\sum_{n} \frac{p_n^2}{p_n^2 + k^2},\label{fvdw2}
\end{equation}
where for convenience we have introduced the notation $\sum'_n$ to denote the sum of terms with $n\neq 0$ but where the $n=0$ term is taken with a weight of $1/2$. A term of the form $\frac{1}{H}\sum'_n $ is a bulk term which is independent of $H$ as the number of modes in a region of length $H$ is $H/\sum'_n 1= a$, where $a$ is a microscopic cut off length scale. The $H$ dependent part of $f_{\rm vdW}$ can thus be written as
\begin{eqnarray}
f^*_{\rm vdW}(H) =- \frac{k_{\mathrm{B}}T}{(2\pi)^2H}\int \mathrm{d}{\bf k}\sideset{}{'}\sum_n \frac{k^2}{p_n^2 + k^2}.\label{fvdw}
\end{eqnarray}
Now we use the identity
\begin{equation}
\sideset{}{'}\sum_n \frac{k^2}{p_n^2 + k^2}=\frac{1}{2}kH\coth(k H)
\end{equation}
to obtain
\begin{equation}
f^*_{\rm vdW}(H)= -\frac{k_{\mathrm{B}}T}{2( 2\pi)^2}\int \mathrm{d}{\bf k}k\coth(k H).
\end{equation}
This term is divergent as $H\to\infty$ but the divergence is independent of $H$, and can therefore be subtracted off to give a $H$ dependent force $f_{\rm vdW}(H)=f^*_{\rm vdW}(H)-\lim_{H\to\infty}f^*_{\rm vdW}(H)$ which is given by
\begin{equation}
f_{\rm vdW}(H)= -\frac{k_{\mathrm{B}}T}{2(2\pi)^2}\int \mathrm{d}{\bf k}k[\coth(k H)-1] = -\frac{k_{\mathrm{B}}T}{4\pi}\int_0^\infty \mathrm{d}kk^2[\coth(k H)-1] .
\end{equation}
The integral over $k$ can then be carried out to find the well known result 
\begin{equation}
f_{\rm vdW}(H)= -\frac{k_{\mathrm{B}}T\zeta(3)}{8\pi H^3},
\end{equation}
where $\zeta(z)$ is the Riemann zeta function. 

\section{Treatment of ions by expanding about mean density}
We now consider the statics and dynamics of the ionic density fluctuations. We write the density of cations/anions  as 
\begin{equation}
\rho_\pm({\bf x}) = \overline \rho +n_\pm({\bf x}).
\end{equation}
Formally we can write the partition function of the ionic part of the system as 
\begin{eqnarray}
Z_{\rm ion} &=& \int \mathrm{d}[\rho_+] \mathrm{d}[\rho_-] \exp\left(-\int \mathrm{d}{\bf x} \left(\rho_+({\bf x})[ \ln(\rho_+({\bf x}))-1]+ \rho_-({\bf x})[ \ln(\rho_-({\bf x}))-1]\right)\right. \nonumber\\
&& ~~~~~~~~~~~~~~~~~~~~~~~~~~~~~~~~~\left. - \frac{\beta q^2}{2}\int \mathrm{d}{\bf x} \mathrm{d}{\bf x}' (\rho_+({\bf x})-\rho_-({\bf x}))G({\bf x},{\bf x}')(\rho_+({\bf x}')-\rho_-({\bf x}'))\right).
\end{eqnarray}
Expanding to quadratic order in the charge fluctuations we find
\begin{equation}
Z_{\rm ion} =\exp(-2V_i~ \overline \rho[\ln\overline\rho-1])\int \mathrm{d}[n_+]\mathrm{d}[n_-]\exp(-{\beta}H_{\rm ion}[n_+,n_-]),
\end{equation}
where the first term corresponds to the partition function of two ideal gases of density $\overline \rho$  in the total  volume $V_i$ between the two slabs corresponding to $z\in[0,H]$. The second quadratic term represents the contribution from the charge fluctuations to the Hamiltonian
\begin{equation}
H_{\rm ion}[n_+,n_-]= \frac{k_{\mathrm{B}}T }{2\overline\rho}\int \mathrm{d}{\bf x} \ [n_+^2({\bf x}) +n_-^2({\bf x})] + \frac{q^2}{2}\int \mathrm{d}{\bf x}\mathrm{d}{\bf x}'\ (n_+({\bf x})-n_-({\bf x}))G({\bf x},{\bf x}')(n_+({\bf x}')-n_-({\bf x}')),
\end{equation}
and the spatial  integration regions above are in the volume $V_i$. Defining the total density fluctuation by
$n_{\mathrm{t}}= n_++n_-$ and the difference $n_{\mathrm{d}} = n_+-n_-$ we can write
\begin{equation}
H_{\rm ion}[n_{\mathrm{t}},n_{\mathrm{d}}]= \frac{k_{\mathrm{B}}T }{4\overline\rho}\int \mathrm{d}{\bf x} \ [n_{\mathrm{t}}^2({\bf x}) +n_{\mathrm{d}}^2({\bf x})] + \frac{q^2}{2}\int \mathrm{d}{\bf x}\mathrm{d}{\bf x}'\ n_{\mathrm{d}}({\bf x})G({\bf x},{\bf x}')n_{\mathrm{d}}({\bf x}')
\end{equation} 
The Gaussian integrals can be performed using the same eigenfunction expansion as in Section (I) and as we consider again the case where $\epsilon\gg \epsilon'$ we find that the $H$ dependent part of the  partition function from each mode is given by
\begin{equation}
Z_{n,\bf{k}}= \left( 1+ \frac{\kappa^2}{k^2+p_n^2}\right)^{-\frac{1}{2}},\label{mode1}
\end{equation}
where $\kappa = \sqrt{2\overline\rho q^2 \beta/\epsilon}$ is the inverse Debye screening length. Computing the force per unit surface area from the resulting free energy then yields
\begin{equation}
f_{\rm ion}=  \frac{k_{\mathrm{B}}T}{(2\pi)^2H}\int \mathrm{d}{\bf k}\sideset{}{'}\sum_{n} \left[\frac{p_n^2}{p_n^2 +\kappa^2+  k^2}- \frac{p_n^2}{p_n^2+  k^2}\right].
\label{screened}
\end{equation}
Notice that the second term cancels exactly with the attractive  van der Waals interaction given in Eq.~(\ref{fvdw2}) and that the first term has exactly the same form but with $k^2$ replaced by $k^2+\kappa^2$. We thus find that the total $H$ dependent force is given by
\begin{equation}
f_{\rm t}(H) = -\frac{k_{\mathrm{B}}T}{4\pi}\int_0^\infty \mathrm{d}k \, k K[\coth( K H)-1] ,
\end{equation}
where $K^2= k^2+ \kappa^2$. Changing the integration to $K$ then gives the (attractive) screened thermal Casimir force
\begin{equation}
f_{\rm t}(H) = -\frac{k_{\mathrm{B}}T}{4\pi}\int_\kappa^\infty \mathrm{d}KK^2[\coth( K H)-1] = -\frac{k_{\mathrm{B}}T}{4\pi H^3}\int_{H\kappa}^\infty \mathrm{d}K \,K^2(\coth K-1) .
\end{equation}
This force is attractive and at large $H$ decays as $\simeq \frac{k_BT \kappa^2}{4\pi  H} \exp(-2\kappa H)$, and hence justifying the terminology ``screened''. We have therefore seen from this calculation that the ionic fluctuations produce a repulsive unscreened Casimir force that exactly cancels that coming from the van der Waals interaction due to the dielectrics, plus an attractive screened interaction that represents the net force.

\section{The stress tensor for the ionic fluctuations}
In order to derive the out of equilibrium forces, one has no free energy from which to derive them and instead one must use a stress tensor to calculate forces. In the presence of an applied electric field, the  density fluctuations give a local body force which is given by~\cite{kru18}
\begin{equation}
{\bf f}_{\rm B}({\bf x}) = -n_+({\bf x}) \nabla\frac{\delta H_{\rm ion}}{\delta n_+({\bf x})}-n_-({\bf x}) \nabla\frac{\delta H_{\rm ion}}{\delta n_-({\bf x})} + q{\bf E}(n_+({\bf x}) - n_-({\bf x})),
\end{equation}
that can be written as
\begin{equation}
{\bf f}_{\rm B}({\bf x}) = -\frac{k_{\mathrm{B}} T}{\overline\rho}n_+({\bf x}) \nabla n_+({\bf x}) -\frac{k_{\mathrm{B}} T}{\overline\rho}n_-({\bf x}) \nabla n_-({\bf x})-q(n_+({\bf x}) -n_-({\bf x}) )\nabla\phi({\bf x}) + q{\bf E}(n_+({\bf x}) - n_-({\bf x})),
\end{equation}
where $\phi({\bf x})$ is the electrostatic field generated by the ionic charges and is given by
\begin{equation}
\nabla\cdot \epsilon({\bf x}) \nabla \phi({\bf x}) = -q (n_+({\bf x}) - n_-({\bf x})).\label{poisson}
\end{equation}
From this it is easy to see that $f_{\mathrm{B}i}({\bf x}) = \nabla_j \sigma_{ij}({\bf x}) + qE_i(n_+({\bf x}) - n_-({\bf x})),$ where $\sigma_{ij}({\bf x})$ is a stress tensor with
\begin{equation}
\sigma_{ij}({\bf x})= -\frac{k_{\mathrm{B}}T}{2\overline \rho}\delta_{ij}(n^2_+({\bf x})+n^2_-({\bf x})) + \epsilon({\bf x})(\nabla_i\phi({\bf r})\nabla_j\phi({\bf x})-\frac{1}{2}\delta_{ij}[\nabla\phi({\bf x})]^2).
\end{equation}
Notice that the second term in $\phi$ is simply the Maxwell stress tensor, while the first two terms correspond to van't Hoff osmotic stresses coming from the density fluctuations.
Noting that  the field ${\bf E}$ is parallel to the surface and that at the surface $\frac{\partial \phi}{\partial z}=0$, we find that the  \emph{outward} force due to the ions per unit surface  acting on the surface at $z=0$ in the direction $z$ is given by
\begin{equation}
f_{\rm ion} = -\langle\sigma_{zz}(0,0,z=0^+) -\sigma_{zz}(0,0,z=0^-)\rangle,
\end{equation}
(there is a change of sign as the inner normal points inwards) which gives
\begin{equation}
f_{\rm ion} = \frac{k_{\mathrm{B}} T}{2\overline \rho}\langle n^2_+({\bf 0})+n^2_-({\bf 0}) \rangle +\frac{\epsilon}{2} \langle [\nabla_{\|}\phi({\bf 0})]^2\rangle,
\label{fistress}
\end{equation}
with ${\bf 0} = (0,0,z=0^+)$ and where we have again invoked the Neumann boundary condition on the field $\phi$, and $\nabla_{\|}$ denotes the gradient in the plane of the surface of the dielectric. Written in terms of the field $n_{\mathrm{t}}$ and $n_{\mathrm{d}}$ this gives Eq.~(11) of the main text
\begin{equation} \label{eq:forceIon}
f_{\rm ion} = \frac{k_{\mathrm{B}} T}{4\overline \rho}\langle n^2_{\rm t}({\bf 0})+n^2_{\rm d}({\bf 0}) \rangle +\frac{\epsilon}{2} \langle [\nabla_{\|}\phi({\bf 0})]^2 \rangle.
\end{equation}
This is the expression that we use in the main text to compute the total force.

\section{Stochastic density functional theory for the driven electrolyte}
The formally exact equation for the stochastic density fields
\begin{equation}
\rho_\pm({\bf x},t) =\sum_{i\pm}\delta({\bf x} - {\bf x}_i(t)) 
\end{equation}
for the cations and anions $i\pm$ in Eq.~(2) of the main text is~\cite{dem16}
\begin{eqnarray}
\frac{\partial \rho_+({\bf x})}{\partial t} = \nabla \cdot D[\nabla \rho_+({\bf x}) + \beta q \rho_+({\bf x}) (\nabla\phi({\bf x}) -{\bf E})] +\nabla\cdot\left( \sqrt{2D\rho_+({\bf x})}~\bm{\eta}_+({\bf x},t)\right),\\
\frac{\partial \rho_-({\bf x})}{\partial t} = \nabla \cdot D[\nabla \rho_-({\bf x}) - \beta q \rho_-({\bf x}) (\nabla\phi({\bf x}) -{\bf E})] +\nabla\cdot \left(\sqrt{2D\rho_-({\bf x})}~\bm{\eta}_-({\bf x},t)\right).
\end{eqnarray}
where the noise terms are Gaussian white noise vector fields in time and space and independent for each species
\begin{equation}
\langle \bm{\eta}_{\mu\alpha}({\bf x},t)\bm{\eta}_{\nu\beta}({\bf x}',t')\rangle = \delta_{\mu\nu}\delta_{\alpha\beta}\delta({\bf x}-{\bf x}')\delta(t-t'), \quad \mu,\,\nu\in\{+,\,-\}.
\end{equation}
The boundary conditions of the fields at the dielectric interfaces are that no current for either species crosses the interface. This no flux boundary condition must be applied to the deterministic and random part of the currents so that
\begin{equation}
{\bf e}_z\cdot [\nabla \rho_\pm({\bf x}) \pm \beta q \rho_\pm({\bf x}) (\nabla\phi({\bf x}) -{\bf E})]= 0\ {\rm and}\ 
{\bf e}_z\cdot \bm{\eta}_\pm=0
\end{equation}
at the surfaces $z=0$ and $z=H$.
These nonlinear stochastic equations are difficult to solve. However if one expands to first order in the fluctuations
\begin{equation}
n_{\pm}({\bf x}) = \rho_\pm({\bf x})-\overline \rho,
\end{equation}
one obtains the soluble linearized SDFT linear equations
\begin{eqnarray}
\frac{\partial n_+({\bf x})}{\partial t} = \nabla \cdot D[\nabla n_+({\bf x}) + \beta q \overline\rho\nabla\phi({\bf x}) -\beta q n_+({\bf x}){\bf E}] +\nabla\cdot \sqrt{2D\overline\rho}~\bm{\eta}_+({\bf x},t),\\
\frac{\partial n_-({\bf x})}{\partial t} = \nabla \cdot D[\nabla n_-({\bf x}) - \beta q \overline\rho \nabla\phi({\bf x}) +\beta q n_-({\bf x}){\bf E}] +\nabla\cdot \sqrt{2D\overline\rho}~\bm{\eta}_-({\bf x},t).
\end{eqnarray}
In terms of the fields $n_{\mathrm{t}}({\bf x})$ and $n_{\mathrm{d}}({\bf x})$ these equations become
\begin{eqnarray}
\frac{\partial n_{\mathrm{t}}({\bf x})}{\partial t} &=& \nabla \cdot D[\nabla n_{\mathrm{t}}({\bf x})  -\beta q n_{\mathrm{d}}({\bf x}){\bf E}] +\nabla\cdot \sqrt{4D\overline\rho}~\bm{\eta}_{\rm t}({\bf x},t),\\
\frac{\partial n_{\mathrm{d}}({\bf x})}{\partial t} &=& \nabla \cdot D[\nabla n_{\mathrm{d}}({\bf x}) +2 \beta q \overline\rho \nabla\phi({\bf x}) -\beta q n_{\mathrm{t}}({\bf x}){\bf E}] +\nabla\cdot \sqrt{4D\overline\rho}~\bm{\eta}_{\rm d}({\bf x},t),\label{end}
\end{eqnarray}
where $\bm{\eta}_{\rm t}({\bf x},t)$ and $\bm{\eta}_{\rm d}({\bf x},t)$ are again independent Gaussian white noise vector fields  in space and time with correlation functions
\begin{equation}
\langle \bm{\eta}_{\mu\alpha}({\bf x},t)\bm{\eta}_{\nu\beta}({\bf x}',t')\rangle = \delta_{\mu\nu}\delta_{\alpha\beta}\delta({\bf x}-{\bf x}')\delta(t-t'), \quad \mu,\,\nu\in\{\mathrm{t},\,\mathrm{d}\}.
\end{equation}
One can use Eq.~(\ref{poisson}) to simplify Eq.~(\ref{end}) which then becomes
\begin{equation}
\frac{\partial n_{\mathrm{d}}({\bf x})}{\partial t} = D(\nabla^2 - \kappa^2 )n_{\mathrm{d}}({\bf x})-D\beta q {\bf E} \cdot \nabla n_{\mathrm{t}}({\bf x})+\nabla\cdot \sqrt{4D\overline\rho}\,\bm{\eta}_{\rm d}({\bf x},t),
\end{equation}
this being the stochastic version of the Debye-Falkenhagen equation. The analysis of the resulting equations is rendered complicated because the no flux boundary condition on the field $n_{\mathrm{d}}({\bf x})$ 
\begin{equation}
\frac{\partial n_{\mathrm{d}}}{\partial z} + 2 \beta q \overline\rho \frac{\partial \phi}{\partial z}=0
\end{equation}
is non local and the  problem of finding the correct eigenfunction expansion  is very involved. However if we consider the limit $\epsilon'\ll\epsilon$ the boundary conditions on the electrostatic field $\phi$ is Neumann and consequently the boundary condition on the field $n_{\mathrm{d}}$ also becomes Neumann. In addition we see that the boundary condition on the field $n_{\mathrm{t}}$ is also Neumann. All of the fields in the problem can thus be written in terms of a Fourier cosine expansion in the $z$ coordinate and a Fourier transform in the $(x,y)$ plane
\begin{eqnarray}
n_{\mathrm{t}}({\bf x}_{\|},z,t) &=&\int \frac{\mathrm{d}{\bf k}}{(2\pi)^2} \sum_n \frac{1}{\sqrt{N_n}}\tilde n_{tn}({\bf k},t)\exp(i{\bf k}\cdot {\bf x}_{\|})\cos(p_n z),\nonumber\\
n_{\mathrm{d}}({\bf x}_{\|},z,t) &=&\int \frac{\mathrm{d}{\bf k}}{(2\pi)^2} \sum_n \frac{1}{\sqrt{N_n}}\tilde n_{dn}({\bf k},t)\exp(i{\bf k}\cdot {\bf x}_{\|})\cos(p_n z)\nonumber,\\
\end{eqnarray}
where $p_n = n\pi/H$ enforces the Neumann conditions at $z=0$ and $z=H$, while $N_n$ normalizes the corresponding eigenfunction, with $N_n = H/2$ for $n\geq 1$ and $N_0=H$. The equations for the Fourier components then read
\begin{eqnarray}
\frac{\partial \tilde{n}_{\mathrm{t}n}({\bf k})}{\partial t} &=& -D(k^2+ p_n^2)\tilde{n}_{\mathrm{t}n}({\bf k})-iD\beta q E k_x \tilde{n}_{\mathrm{d}n}({\bf k}) +\xi_{\mathrm{t}n}({\bf k},t),\\
\frac{\partial \tilde{n}_{\mathrm{d}n}({\bf k})}{\partial t} &=& -D(k^2+ p_n^2+ \kappa^2)\tilde{n}_{\mathrm{d}n}({\bf k})-iD\beta q E k_x \tilde{n}_{\mathrm{t}n}({\bf k}) +\xi_{\mathrm{d}n}({\bf k},t),
\end{eqnarray}
where, $k_x={\bf e}_x\cdot{\bf k}$, the noise terms  $\xi_{\mathrm{t}n}({\bf k},t)$ and $\xi_{\mathrm{d}n}({\bf k},t)$ are independent with correlation functions
\begin{equation}
\langle \xi_{\mu n}({\bf k},t)\xi_{\nu n'}({\bf k}',t')\rangle = (2\pi)^2\delta(t-t')\delta({\bf k}+{\bf k}')\delta_{\mu\nu} \delta_{nn'}~4D\overline\rho(k^2+ p_n^2), \quad \mu,\,\nu\in\{\mathrm{t},\,\mathrm{d}\}.
\end{equation}
The equal time correlation functions of the Fourier components in the steady state are denoted as
\begin{equation}
\langle \tilde{n}_{\mu n}({\bf k})\tilde{n}_{\nu n}({\bf k}')\rangle = (2\pi)^2\delta({\bf k}+{\bf k'})~\tilde{C}_{\mu\nu n}({\bf k}), \quad \mu,\, \nu \in \{\mathrm{t},\,\mathrm{d}\}
\end{equation}
and we can define the correlation matrix 
\begin{equation}
C_n({\bf k}) = \begin{pmatrix} \tilde{C}_{\mathrm{tt}n}({\bf k}) & \tilde{C}_{\mathrm{td}n}({\bf k})\\ \tilde{C}_{\mathrm{dt}n}({\bf k}) & \tilde{C}_{\mathrm{dd}n}({\bf k})\end{pmatrix}.
\end{equation}
It is now standard to see that in the steady state that  $C_n({\bf k})$ obeys the so called Lyapunov equation
\begin{equation}
A_n({\bf k}) C_n({\bf k})+ C_n({\bf k})A_n(-{\bf k})= 2 R_n({\bf k}),
\end{equation}
where 
\begin{equation}
A_n({\bf k})=D\begin{pmatrix} k^2+ p_n^2 & \mathrm{i}k_x \beta q E\\ \mathrm{i}k_x \beta q E & k^2 + p_n^2 + \kappa^2 \end{pmatrix},
\end{equation}
and 
\begin{equation}
R_n({\bf k})=2D\overline{\rho} \begin{pmatrix} k^2+ p_n^2 & 0\\ 0 & k^2 + p_n^2 \end{pmatrix}.
\end{equation}
Solving  this linear system then gives
\begin{eqnarray}
\tilde{C}_{\mathrm{tt}n}({\bf k}) &=& 2\overline{\rho} \frac{(k^2 + p_n^2) \left[ (K^2 + p_n^2) (K^2 + k^2 + 2 p_n^2) + 2 \beta^2 q^2 E^2 k_x^2  \right]}{(K^2 + k^2 + 2 p_n^2)\left[(K^2 + p_n^2)(k^2 + p_n^2) + \beta^2 q^2 E^2 k_x^2  \right]}\label{ctt}, \\
\tilde{C}_{\mathrm{td}n}({\bf k}) &=&-\tilde{C}_{\mathrm{dt}n}({\bf k}) = 2\mathrm{i}\overline{\rho} \frac{\beta q E k_x \kappa^2 (k^2 + p_n^2)}{(K^2 + k^2 + 2 p_n^2)\left[(K^2 + p_n^2)(k^2 + p_n^2) + \beta^2 q^2 E^2 k_x^2  \right]}, \\
\tilde{C}_{\mathrm{dd}n}({\bf k}) &=&  2\overline{\rho} \frac{(k^2 + p_n^2) \left[ (k^2 + p_n^2) (K^2 + k^2 + 2 p_n^2) + 2 \beta^2 q^2 E^2 k_x^2  \right]}{(K^2 + k^2 + 2 p_n^2)\left[(K^2 + p_n^2)(k^2 + p_n^2) + \beta^2 q^2 E^2 k_x^2  \right]}\label{cdd}.
\end{eqnarray}
This is  the solution of the Lyapunov equation that we use in the main text.

\section{Computation of the force due to the ions}
Using the Fourier representation of the correlation functions and Eq.~(\ref{eq:forceIon}) for the ionic force in terms of the stress tensor we find that
\begin{equation}
f_{\rm ion}= \frac{k_\mathrm{B} T}{H}\int \frac{\mathrm{d}{\bf k}}{(2\pi)^2}\sideset{}{'}\sum_n S_n({\bf k}),
\end{equation}
where 
\begin{equation}
S_n({\bf k})= \frac{1}{2\overline \rho}(\tilde{C}_{\mathrm{tt}n}({\bf k} )+ \tilde{C}_{\mathrm{dd}n}({\bf k}) ) +\frac{\beta q^2k^2 \tilde{C}_{\mathrm{dd} n}({\bf k}) }{\epsilon(k^2 +p_n^2)^2}
= \frac{1}{2\overline{\rho}}\left[\tilde{C}_{\mathrm{tt}n}({\bf k} )+ \tilde{C}_{\mathrm{dd}n}({\bf k}) + \frac{\kappa^2k^2 \tilde{C}_{\mathrm{dd}n}({\bf k}) }{(k^2 +p_n^2)^2}\right].
\end{equation}
Inserting the expressions in Eq.~(\ref{ctt}) and Eq.~(\ref{cdd}) then gives
\begin{equation}
S_n({\bf k})= 2+ \frac{2k^2}{k^2+ p_n^2}- \frac{2K^2 + 2k^2}{K^2 + k^2 + 2 p_n^2}
+ \frac{p_n^2 \kappa^4}{( K^2 +k^2+ 2 p_n^2)\left((K^2 + p_n^2)(k^2 + p_n^2) + \beta^2\ q^2 E^2 k_x^2\right)} .
\end{equation}
Before examining the effect of driving we shall consider the case $E=0$ which gives
\begin{equation}
S_n({\bf k})= 2+ \frac{k^2}{k^2+ p_n^2}-\frac{k^2+\kappa^2}{k^2+ \kappa^2+ p_n^2}.
\end{equation}
Using this it is straightforward to verify that we recover that the resulting $H$ dependent force is in perfect agreement with the static result as given in Eq.~(\ref{screened}).

As we are interested in the total force $f_{\rm t}= f_{\rm ion} + f_{\rm vdW}$, we will add on the van der Waals contribution to the effective interaction which can be read off from Eq.~(\ref{fvdw}) and write
\begin{equation}
f_{\rm t}= \frac{k_{\mathrm{B}}T}{H}\int \frac{\mathrm{d}{\bf k}}{(2\pi)^2} \sideset{}{'}\sum_n S_{\mathrm{t}n}({\bf k}),
\end{equation}
where 
\begin{equation}
S_{\mathrm{t}n}({\bf k})= S_n({\bf k})-\frac{k^2}{p_n^2 + k^2}.
\end{equation}
This  gives
\begin{equation}
S_{\mathrm{t}n}({\bf k})= 2+\frac{k^2}{k^2+ p_n^2} -2\frac{K^2 + k^2}{K^2 + k^2 + 2 p_n^2}
+ \frac{p_n^2 \kappa^4}{( K^2 +k^2+ 2 p_n^2)\left((K^2 + p_n^2)(k^2 + p_n^2) + \beta^2\ q^2 E^2 k_x^2\right)} .
\end{equation}
The long range component of the force can be deduced from the low wave vector behavior of $S_{\mathrm{t}n}({\bf k})$ which is given by
\begin{equation}
S_{\mathrm{t}n}({\bf k})\approx 2+
 \frac{p_n^2}{k^2 + p_n^2 + \frac{\beta^2\ q^2 E^2 k_x^2}{\kappa^2}} +\frac{k^2}{k^2+ p_n^2},
\end{equation}
which has a long range component (the constant term $2$ yields a bulk term independent of $H$)
\begin{equation}
S_{tn}({\bf k})\approx  \frac{k^2}{k^2 + p_n^2}-\frac{Q^2}{p_n^2 + Q^2} ,
\end{equation}
where
\begin{equation}
Q = \sqrt{1+\frac{\beta^2 E^2 q^2 \cos^2\theta}{\kappa^2}} k
\end{equation}
and we have written $k_x=k\cos\theta$.
Now computing the $H$ dependent part of the interaction in exactly the same way as the similar calculations performed above we find
\begin{equation}
f_{\rm t}(H) \approx -\frac{1}{8\pi^2\beta}\int \mathrm{d}{\bf k} [Q(\coth(QH)-1)-k(\coth(kH)-1)].
\end{equation}
The remaining integrals can be carried out to yield
\begin{equation}
f_{\rm t}(H)\approx \frac{\zeta(3)}{8\pi\beta H^3}\left[1-\frac{1}{\sqrt{1+\frac{\beta^2 q^2E^2 }{\kappa^2}}}\right].\label{bigH}
\end{equation}
We now discuss the case where $H$ is finite but the large driving limit $E\to\infty$ is taken. Here we can write
\begin{equation}
S_{tn}({\bf k})= 2+\frac{k^2}{k^2+ p_n^2} -2\frac{k^2+ \frac{\kappa^2}{2}}{ k^2 + p_n^2 + \frac{\kappa^2}{2}},
\end{equation}
from which we find that the $H$ dependent force is given by
\begin{equation}
f_{\rm t}(H, E\to\infty) = \frac{\zeta(3)}{8\pi\beta H^3} - \frac{2}{4\pi\beta H^3}\int_{\frac{H\kappa}{\sqrt{2}}}^\infty \mathrm{d} K K^2(\coth K-1).\label{strong}
\end{equation}
The first term is the repulsive long range Casimir force seen in Eq.~(\ref{bigH}) and the second term is twice the screened thermal Casimir interaction for an electrolyte with inverse screening length $\kappa^*= \kappa/\sqrt{2}$. 
This screening length corresponds to $\kappa^* = \sqrt{\overline\rho q^2 \beta/\epsilon}$, that is to say the density $\overline\rho$ is replaced by $\overline\rho/2$. 
This halving of the effective density and the appearance of a factor of $2$ in the screened contribution can be explained by invoking a decoupling mechanism between the cations and anions.
The field drives the cations and anions with velocities $v_\pm=
\pm \beta D q E$.
Each species moves with corresponding average velocity in its own Galilean reference frame where  they equilibrate with each other. However, the strong driving means that  the interactions between the different species effectively disappear, the cations and anions thus behave as if they are in equilibrium but  are decoupled. The Hamiltonian  for the charge fluctuations in this strongly driven limit is thus
\begin{equation}
H_{\rm decoupled}[n_+,n_-]= \frac{k_{\mathrm{B}}T }{2\overline{\rho}}\int \mathrm{d}{\bf x} [\ n_+^2({\bf x}) +n_-^2({\bf x})] + \frac{q^2}{2}\int \mathrm{d}{\bf x}\mathrm{d}{\bf x}' \left[\ n_+({\bf x})G({\bf x},{\bf x}')n_+({\bf x}')+n_-({\bf x})G({\bf x},{\bf x}')n_-({\bf x}')\right],
\end{equation}
that is to say we have removed the interactions between charges of different species.

The partition function is thus given by
\begin{equation}
Z_{\rm decoupled}= Z_1^2,
\end{equation}
where
\begin{equation}
Z_1 = \int \mathrm{d}[n] \exp\left(-\frac{1}{2\overline{\rho}}\int \mathrm{d}{\bf x} \ n^2({\bf x}) - \frac{\beta q^2}{2}\int \mathrm{d}{\bf x}\mathrm{d}{\bf x}'\ n({\bf x})G({\bf x},{\bf x}')n({\bf x}')\right)
\end{equation}
is the partition function for a single species. Computing $Z_1$ gives a contribution from each mode of 
\begin{equation}
Z_{1,n,\bf{k}}= \left( 1+ \frac{\frac{\kappa^2}{2}}{k^2+p_n^2}\right)^{-\frac{1}{2}}.
\end{equation}
Comparing with the zero driving equilibrium computation from Eq.~(\ref{mode1}) then gives the force due to the decoupled ions as 
\begin{equation}
f_{\rm decoupled}= 2\times \frac{k_{\mathrm{B}}T}{(2\pi)^2H}\int \mathrm{d}{\bf k}\sideset{}{'}\sum_{n} \left[\frac{p_n^2}{p_n^2 +\frac{\kappa^2}{2}+  k^2}- \frac{p_n^2}{p_n^2+  k^2}\right].
\end{equation}
Computing this and then adding the attractive van der Waals component gives exactly the strong driving  nonequilibrium result Eq.~(\ref{strong}).


\begin{thebibliography}{999}
\bibitem{woo16}L.M. Woods, D.A.R. Dalvit, A. Tkatchenko, P. Rodriguez-Lopez, A.W. Rodriguez  and R. Podgornik, Materials perspective on Casimir and van der Waals interactions, Rev. Mod. Phys. {\bf 88}, 045003 (2016).
\bibitem{dan23}D.M. Dantchev and S. Dietrich, Critical Casimir Effect: Exact Results, Phys. Rep., {\bf 1005}, 1 (2023).
\bibitem{par05} V.A. Parsegian, Van der Waals Forces: A Handbook for Biologists, Chemists, Engineers, and Physicists, Cambridge University Press (Cambridge) (2005).
\bibitem{RevModPhys.90.045001} {A. Macio\l{}ek and S. Dietrich}, Collective behavior of colloids due to critical Casimir interactions, {Rev. Mod. Phys.}, {\bf 90}, {045001} (2018).\bibitem{dea16}D. S. Dean, B.-S. Lu, A. C. Maggs, and R. Podgornik, Nonequilibrium Tuning of the Thermal Casimir Effect, Phys. Rev. Lett. {\bf 116}, 240602 (2016).
\bibitem{mah21} S. Mahdisoltani and R. Golestanian, Long-Range Fluctuation-Induced Forces in Driven Electrolytes, Phys. Rev. Lett. {\bf 126}, 158002 (2021).
\bibitem{Advances} M. Bordag, G. L. Klimchitskaya, U. Mohideen, and V. M. Mostepanenko, Advances in the Casimir Effect (Oxford University Press, New York, 2009).
\bibitem{Antezza2} M. Antezza, L.P. Pitaevskii, S. Stringari, V.B. Svetovoy, Casimir-Lifshitz force out of thermal equilibrium and asymptotic nonadditivity, Phys. Rev. Lett. {\bf 97}, 223203 (2006). 
\bibitem{Antezza1} M. Antezza, L. P. Pitaevskii, S. Stringari, and V. B. Svetovoy, Casimir-Lifshitz force out of thermal equilibrium, Phys. Rev. A {\bf 77}, 022901 (2008).
\bibitem{Kardar} M. Kr\" uger, T. Emig, and M. Kardar, Nonequilibrium Electromagnetic Fluctuations: Heat Transfer and Interactions, Phys. Rev. Lett. {\bf 106}, 210404 (2011).
\bibitem{dea12} D. S. Dean, V. D\' emery, V. A. Parsegian, and R.  Podgornik, Out-of-equilibrium relaxation of the thermal Casimir effect in a model polarizable material, Phys. Rev. E {\bf 85}, 031108 (2012).
\bibitem{dea14} D. S. Dean and R. Podgornik,  Relaxation of the thermal Casimir force between net neutral plates containing Brownian charges, Phys. Rev. E {\bf 89}, 032117 (2014).
\bibitem{lu15} B.-S. Lu, D. S. Dean and R. Podgornik, Out of equilibrium thermal Casimir effect between Brownian conducting plates, Europhys. Lett., {\bf 112}, 20001,  (2015).
\bibitem{lev99} Y. Levin, When Do Like Charges Attract?, Physica A 265, 432 (1999).
\bibitem{jan05}B. Jancovici and L. Samaj, Casimir force between two ideal-conductor walls revisited, Europhys. Lett. {\bf 72}, 35 (2005)



\bibitem{nin76} B.W. Ninham and J. Mahanty, Dispersion forces, London Academic Press (1976).
\bibitem{net01} R.R. Netz, Static van der Waals interactions in electrolytes, Eur. Phys. J. E {\bf 5}, 189 (2001).


\bibitem{gar90}P. L. Garrido, J. L. Lebowitz, C. Maes, and H. Spohn, Long-range correlations for conservative dynamics, Phys. Rev. A
{\bf 42}, 1954 (1990).
\bibitem{gri90}G. Grinstein, D.-H. Lee, and S. Sachdev, Conservation Laws, Anisotropy, and Self-Organized Criticality in Noisy Non-
equilibrium Systems, Phys. Rev. Lett. {\bf 64}, 1927 (1990).
\bibitem{Sengers1} T. R. Kirkpatrick, J. M. Ortiz de Z\' arate, and J. V. Sengers, Giant Casimir Effect in Fluids in Nonequilibrium Steady States,  Phys. Rev. Lett. {\bf 110}, 235902 (2013).
\bibitem{Sengers2} T. R. Kirkpatrick, J. M. Ortiz de Z\' arate, and J. V. Sengers, Fluctuation-induced pressures in fluids in thermal nonequilibrium steady states, Phys. Rev. E {\bf 89}, 022145 (2014).
\bibitem{Perkin2019} C. S. Perez-Martinez and S. Perkin, Surface forces generated by the action of electric fields across liquid films, Soft Matter {\bf 15} 4255 (2019).
\bibitem{dem16}V. D\'emery and D. S. Dean, The conductivity of strong electrolytes from stochastic density functional theory, J. Stat. Mech.  023106 (2016).
\bibitem{dea96}D. S. Dean, Langevin equation for the density of a system of interacting Langevin processes, J. Phys. A {\bf 29}, L613 (1996).
\bibitem{SM} See {S}upplementary {M}aterial for more detailed explanation of the calculation.
\bibitem{Future} G. Du, D.S. Dean, B. Miao,  and R. Podgornik, in preparation.
\bibitem{kru18}M. Kr\"uger, A. Solon, V. D\'emery, C. M. Rohwer, and D. S. Dean, Stresses in non-equilibrium fluids: Exact formulation and coarse-grained theory, J. Chem. Phys. {\bf 148}, 084503 (2018).

\bibitem{jan04}B. Jancovici and L. Šamaj, Screening of classical Casimir
forces by electrolytes in semi-infinite geometries, J. Stat. Mech. P08006 (2004).
\bibitem{att87} P. Attard, R. Kjellander and D. J. Mitchell, Interactions between electroneutral surfaces nearing mobile charges, Chem. Phys. Lett. {\bf 139}, 219 (1987).
\bibitem{att88} P. Attard, D. J. Mitchell, and B. W. Ninham, Beyond Poisson–Boltzmann: Images and correlations in the electric double layer. I. Counterions only, J. Chem. Phys. {\bf 88}, 4987 (1988).
\bibitem{dea02}D.S. Dean and R.R. Horgan, Electrostatic fluctuations in soap films, Phys. Rev. E {\bf 65}, 061603 (2002).
\bibitem{ric20} L. Richter, P. J. Żuk, P. Szymczak, J. Paczesny, K. M. Bkak,
T. Szymborski, P. Garstecki, H. A. Stone, R. Hołyst, and C.Drummond, Ions in an AC Electric Field: Strong Long- Range Repulsion Between Oppositely Charged Surfaces, Phys. Rev. Lett. {\bf 125}, 056001 (2020).
\bibitem{Trefalt20} A. Smith, M. Borkovec, G.  Trefalt, Forces between solid surfaces in aqueous
electrolyte solutions,  Adv. Colloid Interface Sci. {\bf  275}, 102078 (2020).
\end{thebibliography}

\begin{thebibliography}{999}
\bibitem{jan04}B. Jancovici and L. Šamaj, Screening of classical Casimir
forces by electrolytes in semi-infinite geometries, J. Stat. Mech. P08006 (2004).
\bibitem{kru18}M. Kr\"uger, A. Solon, V. D\'emery, C. M. Rohwer, and D. S. Dean, Stresses in non-equilibrium fluids: Exact formulation and coarse-grained theory, J. Chem. Phys. {\bf 148}, 084503 (2018).
\bibitem{dem16}V. D\'emery and D. S. Dean, The conductivity of strong electrolytes from stochastic density functional theory, J. Stat. Mech.  023106 (2016).
\end{thebibliography}
\end{document}